\documentclass[graybox]{svmult}

\usepackage{type1cm}        
\usepackage{makeidx}         
\usepackage{graphicx}        
\usepackage{multicol}        
\usepackage[bottom]{footmisc}

\usepackage{newtxtext}       %
\usepackage{newtxmath}       

\usepackage{soul}
\usepackage{todonotes}
\usepackage{xcolor}

\makeindex             

\begin{document}

\title*{Extended Reality for Knowledge Work in Everyday Environments}
\author{Verena Biener, Eyal Ofek, Michel Pahud, Per Ola Kristensson and Jens Grubert}
\institute{
Author's version. \\ \\
Verena Biener \at Coburg University of Applied Sciences and Arts, \email{verena.biener@hs-coburg.de}
\and Eyal Ofek \at Microsoft Research Redmond \email{eyalofek@microsoft.com}
\and Michel Pahud \at Microsoft Research Redmond \email{mpahud@microsoft.com}
\and Per Ola Kristensson \at University of Cambridge \email{pok21@cam.ac.uk}
\and Jens Grubert \at Coburg University of Applied Sciences and Arts, \email{jens.grubert@hs-coburg.de}
}
%
%
\maketitle

\abstract{  Virtual and Augmented Reality ({\em VR} and {\em AR}) have the potential to change information work. The ability to modify the workers senses can transform everyday environments into a productive office, using portable head-mounted displays ({\em HMDs})  combined with conventional interaction devices, such as keyboards and tablets. While a stream of better, cheaper and lighter HMDs have been introduced for consumers in recent years, there are still many challenges to be addressed to allow this vision to become reality. This chapter summarizes the state of the art in the field of extended reality for knowledge work in everyday environments and proposes steps to address the open challenges.} 

\section{Introduction}

Extended Reality (XR) covers a spectrum of diverse technologies ranging from augmented physical  ({\em AR}) environments to fully virtual ({\em VR}) environments. While XR technologies have been studied for decades in laboratory settings, the shift towards affordable consumer-oriented products allows exploring the positive and negative qualities of such technologies in everyday environments. Among diverse application domains, spanning entertainment, medical and industrial use \cite{slater2016enhancing, billinghurst2015survey}, supporting knowledge work has attracted increasing interest in recent years.

The notation {\em 'Knowledge Work'} follows the definition initially coined by Peter Drucker \cite{drucker1966effective} where information workers (or {\em IW}s) apply theoretical and analytical knowledge to develop products and services. Much of the work might be detached from physical documents, artifacts or specific work locations, and is mediated through digital devices such as laptops, tablets or mobile phones, connected through the internet. This unique nature of knowledge work enables better mobility---the ability to work far from the physical office---and raises new possibilities to both overcome difficulties and enable services that are natural for physical offices, such as streamlined communication and collaboration and environments designed for creativity, privacy and other factors. 

In this chapter, we hope to explain that moving the worker further into a digital immersive environment opens up further options not limited by the physicality of the worker's devices and environments. We hypothesize that a virtual work environment enables workers to do more than they could before, with less effort, and may also eventually level the plain-field between workers regardless of their physical locations, physical limitations, or the quality of their physical work environment. We summarize both existing research on supporting knowledge work through XR technologies and challenges that should be addressed to move the field forward.

The focus of this work is to explore research in the area of extended reality that takes place in a knowledge work context.

Our search includes diverse papers from journals and conferences and is not limited to certain venues. Most papers were found through ACM Digital Library, IEEE Explore and Google Scholar.
The following search terms were used in different combinations: virtual reality, augmented reality, knowledge work, text entry, collaboration, office environment, and long term use. These were used in the advanced search engines of aforementioned databases. Additionally, we considered papers that referenced, or were referenced, by relevant papers. We primarily examined the title and abstract to decide, based on our own expertise in the field, whether a paper was relevant. We used the following criteria: 1) the paper included an XR technology; and 2) the paper had a connection to the field of knowledge work. 



This chapter is divided into several sections, each investigating a different aspect of knowledge work in XR. Section \ref{Section:interaction}, {\em Interaction Techniques}, reviews research on interaction techniques that can facilitate knowledge work in XR, including techniques for text entry (\ref{TextEntry})---a crucial task in knowledge work. Section \ref{Section:collaboration}, {\em Collaboration}, explores the collaborative aspects of information work, which are crucial to allow the worker to maintain group working from remote and new environments. Section \ref{Section:Environments}, {\em Environment}, considers the influence of the environment on the knowledge worker and how XR can help optimize it, as well as the social implications arising from the use of XR. Section \ref{Section:Applications}, {\em Application}, presents practical applications of XR in the area of knowledge work. Finally, we envision  the worker to use the XR environment as an alternative to the limiting physical environment, resulting in the use of XR for extended time periods. Therefore Section \ref{LOngTerm}, {\em Long-term Immersion}, provides insights into the current research status in this area. Towards the end of this chapter we provide a summary, discuss future challenges, and synthesize our main conclusions.

\section{Interaction Techniques}
\label{Section:interaction}

Envisioning work in XR spaces, we look at ways that information workers are doing their task today, and how they can be done in XR space. The transition to XR space, where the user's senses are being modified by devices such as Head Mounted Displays (HMDs) and different interfaces for the hands are challenging. Some tasks such as typing which are crucial in today's work, are dependent on good combination of senses (such as vision, proprioception and haptics). 

\begin{figure}[b]
	\centering 
	\includegraphics[width=0.6\columnwidth]{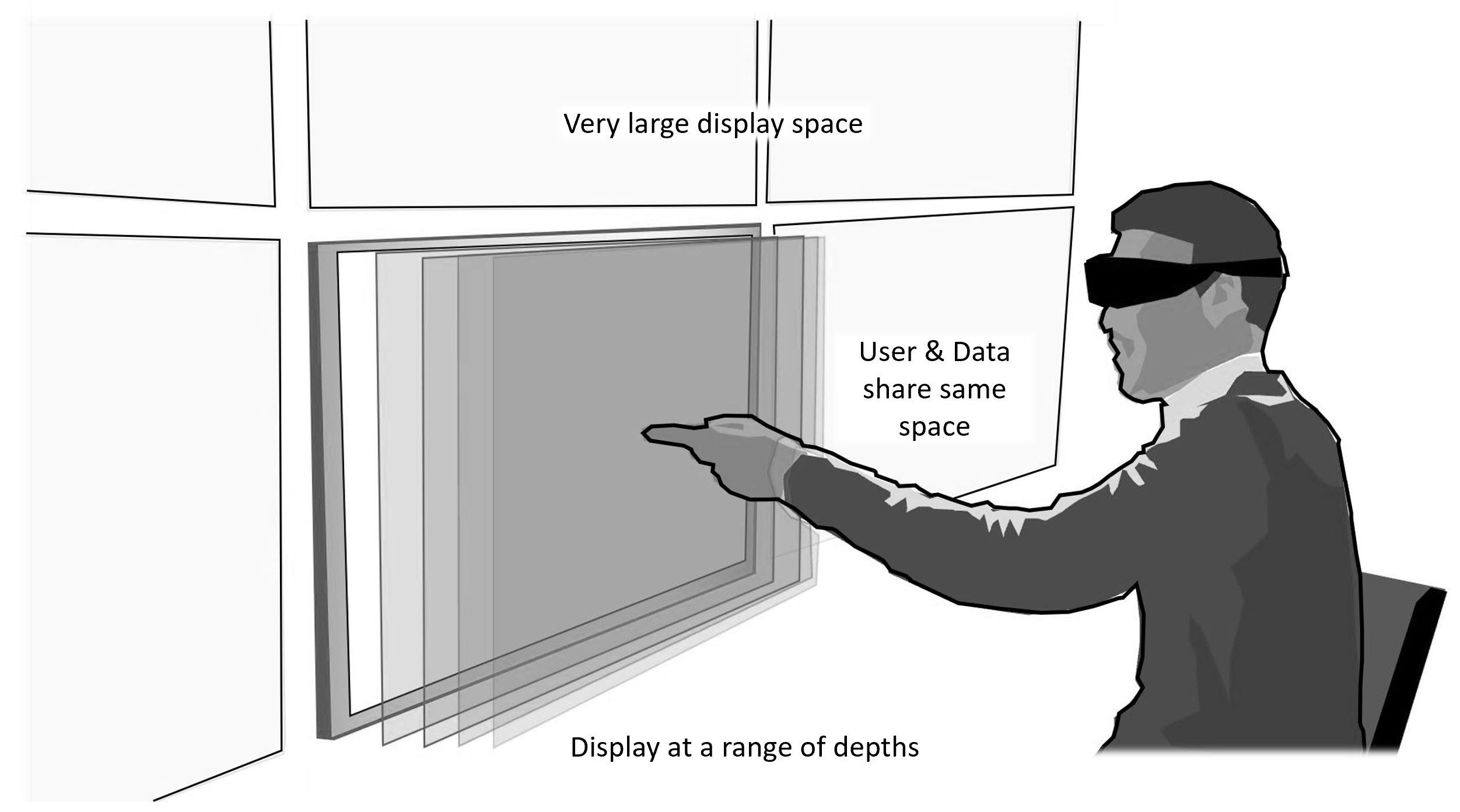}
	\caption{Using a head mounted display adds several capabilities over conventional 2D displays. First, the effective display field of view engulfs the user, second the HMD's stereo display  enables the position of data at different distances from the user, and is not limited to a planar display. Finally, positioning the user and the data in the same space enables new and direct ways for the user to interact with the data using natural gestures.}
	\label{fig:MRAddAxeses}
\end{figure}

Another aspect of interest is embedding the, mostly 2D, information work in three-dimensional space. The use of a 3D immersive space enables several capabilities that were not available to workers using standard 2D monitors and input devices, such as a very wide display space around the user, a depth display that is not limited to one plane as most monitors, and storing data in the same space as the user's body, enabling new and direct methods to interact with data using natural gestures (Figure \ref{fig:MRAddAxeses}).
Different works look at how to efficiently map user's 3D motions in space to the 2D space of the task or documents (e.g. spreadsheets or letters), while other looked at how to use the additional dimension of display given by XR to expose more meta data about the task that can help the worker. 

\subsection{Working with 2D Content in 3D}
Toady's most common knowledge worker tasks, such as document editing or spreadsheet applications, are done in 2D on  2D displays. Therefore, in order to leverage on the existing data, tools and user's familiarity, many approaches that combine knowledge work with XR three-dimensional environments keeps the 2D nature of such tasks and develop techniques for interacting with 2D content in a 3D environment.

\begin{figure}[t]
	\centering 
	\includegraphics[width=0.8\columnwidth]{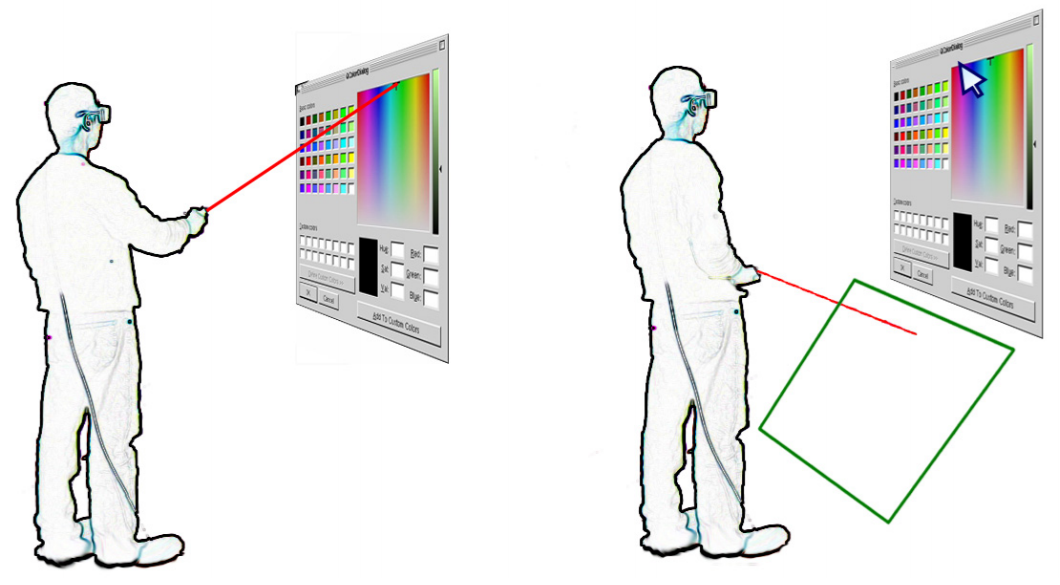}
	\caption{Virtual pad as proposed by Andujar and Argelaguet \cite{andujar2007virtual}. On the left, the user directly interacts with the 2D content through ray-casting. On the right, the user casts the ray onto a virtual pad which redirects the movement to the 2D content and allows a more comfortable position. Image courtesy by Ferran Argelaguet.}
	\label{fig:virtualPads}
\end{figure}

Immersive 3D environments offer a large space around the user that can be used to display more information. However, directly interacting with the content, for example through gestures, can be fatiguing, especially for large displays. This problem can be tackled by indirect interaction.
For example, Andujar and Argelaguet \cite{andujar2007virtual} proposed to interact with 2D windows inside virtual reality by decoupling motor space from visual space. To this end, users interacted with a controller pointing on a virtual pad, which mapped movements to the respective 2D window (see Figure \ref{fig:virtualPads}). They compared  the technique with direct manipulation ray-casting and the performance results indicated that there is a small overhead. However, authors argue that it is a good trade-off for more flexibility and comfort.
Later, this idea was also studied within AR \cite{brasier2020arpads} where results suggested that indirect input can perform equally to direct hand raycast and produces less fatigue. 
Also, Hoppe et al. \cite{hoppe2020enabling}, proposed a set of tools to further interact with 2D Window content, such as a workbench tool for copying and pasting 2D content within 3D or a macro tool, inside 3D virtual environments.

Norman et al. \cite{normand2018enlarging} used augmented reality to extend the display space of a smartphone and additionally presented new mid-air interaction techniques. The results of their study indicate that the extended display space was superior to using the smartphone alone. Additionally, input on the phone performed better than the proposed mid-air interaction techniques.  Le et al. \cite{le2021vxslate} presented VXSlate, combining head tracking and a tablet to perform fine-tuned manipulations on a large virtual display. A virtual representation of the users hand and the tablet is presented on the large display. Kern et al. \cite{kern2021off} introduced a framework for 2D interaction in 3D including digital pen and paper on physically aligned surfaces and a study (n=10) showed that the technique resulted in low task load and high usability.

\begin{figure}[t]
	\centering 
	\includegraphics[width=0.8\columnwidth]{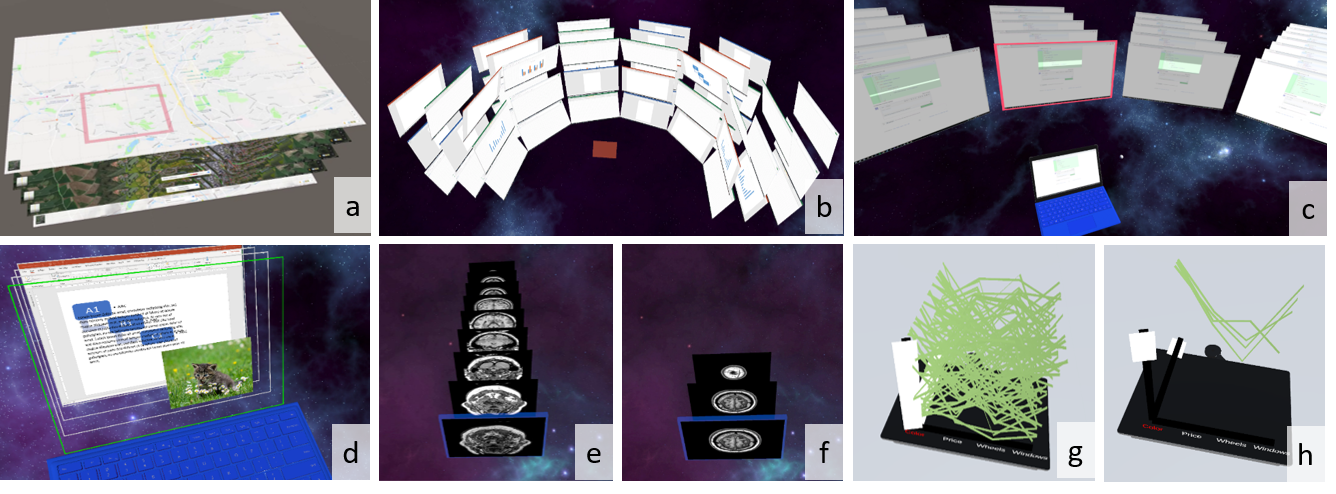}
	\caption{Applications proposed by Biener et al. \cite{biener2020breaking} using the joint interaction space of tablets and immersive VR HMDs for tasks like map navigation (a), window manager (b), code version control (c), presentation editor (d), medical imaging (e,f) and information visualization (g,h)}
	\label{fig:breakingTheScreen}
\end{figure}

Biener et al. \cite{biener2020breaking} investigated the joint interaction space of tablets and immersive VR HMDs for supporting the interaction of knowledge workers with and across multiple 2D windows embedded in a 3D space (see Figure \ref{fig:breakingTheScreen}). Specifically, they designed techniques, including touch and eye-gaze, and therefore combined indirect and direct techniques. The goal was to be unobtrusive and compatible with small physical spaces within everyday environments like trains and planes.
Using physical devices like tablets or pens can also minimizes motions and provides hand support for long hours of work and therefore reduce fatigue.
For example, Gesslein et al. \cite{gesslein2020pen} proposed pen and gaze-based interaction techniques for supporting interaction with spreadsheets in VR. Aside from the large display space they also made use of the 3D view to show additional information above the 2D screen.
Einsfeld et al. \cite{einsfeld2006dynamic} also presented a semantic information visualization of documents in a 3D interface, which visualizes documents, meta-data and semantic relations between documents. 
Dengel et al. \cite{dengel2006human} extended this work through interaction techniques for searching and navigating large document sets with a stereoscopic monitor and a data glove. Evaluating these techniques indicated that they are fun and enable an efficient interaction.
Further, Deller et al. \cite{deller2008managing} introduced interaction techniques for managing 2D documents within a virtual work desk.



As has been seen, many proposed techniques include a touch surface \cite{normand2018enlarging, biener2020breaking, gesslein2020pen, le2021vxslate}, which enables fine grained sensing of writing and touch, while supporting the users fingers or stylus, and  has found to perform better than mid-air techniques \cite{normand2018enlarging,gesslein2020pen, romat2021flashpen, kern2021off}. 
The large display space of XR enable display of more data than typical physical displays, yet the need to manipulate data over such large space efficiently without generating fatigue of the user is a challenge. Indirect input has been shown to performs equally to direct while inducing less fatigue \cite{brasier2020arpads}. The user may execute smaller gestures, and may support their hands to enable long hours of work, and maps their action to the large display space. This is not a new concept for information workers, used for indirect mapping of mouse inputs.

\subsection{System Control}
\label{Section:sysControl}

While many techniques have been proposed for task such as object manipulation (for a recent survey we refer to \cite{mendes2019survey}), research on system control has not been as much in the focus of attention. Within the context of seated Virtual Reality, Zielasko et al. \cite{zielasko2019passive}, studied the effects of passive haptic feedback on touch-based menus. They found a mid-air menu with passive haptic-feedback to outperform desk-aligned alternatives (with and without haptic feedback). However, they also noted hardware requirements for supporting passive haptic feedback in VR, which might be challenging to achieve in everyday environments.  
Bowman et al. \cite{bowman2001design} compared a menu system using Pinch Gloves (TULIP) with floating menus and pen and tablet menus. They found that users had a preference for TULIP although pen and tablet was significantly faster. This was due to the TULIP interface providing less affordances.

\begin{figure}[t]
	\centering 
	\includegraphics[width=0.8\columnwidth]{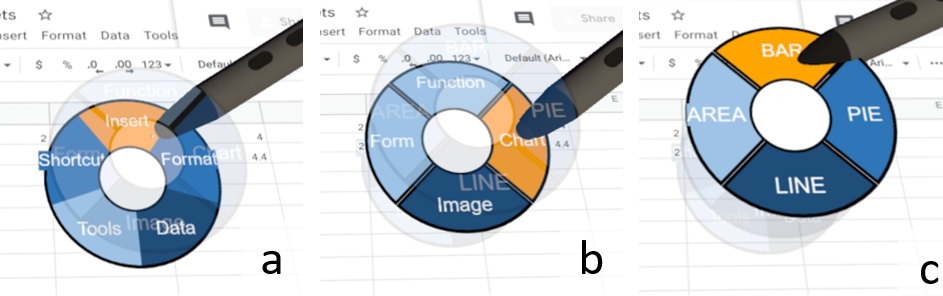}
	\caption{Stacked radial menu in the spreadsheet application presented in \cite{gesslein2020pen}. }
	\label{fig:radialStackedMenu-spreadsheet}
\end{figure}


\begin{figure}
	\centering 
	\includegraphics[width=0.5\columnwidth]{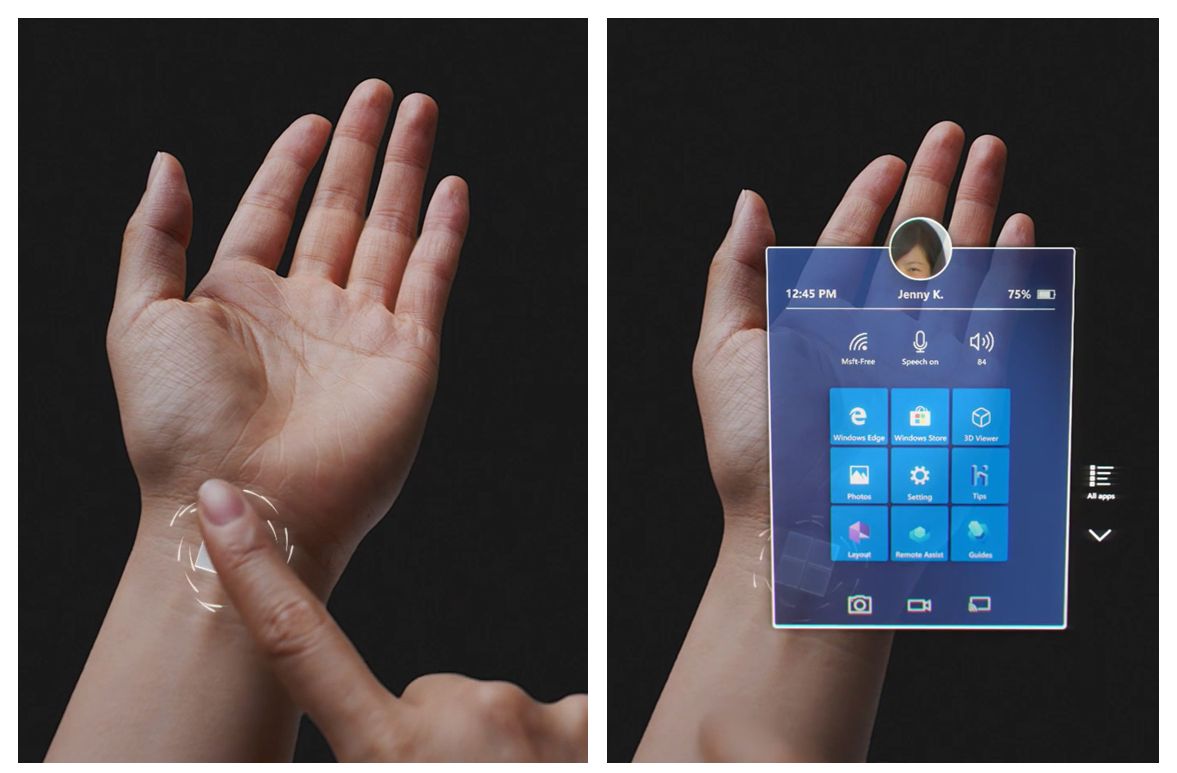}
	\caption{Touching a wrist with a finger is used to summon a menu in HoloLens2 HMD. }
	\label{fig:HL2Start}
\end{figure}

In current HMDs system control is usually realized via gestures, controllers and pointing. For example, the Oculus Quest opens a rectangular menu-window upon pressing the menu-buttons on the controller, then the user can navigate through the menu using raycasting and the controllers trigger. Alternatively, a hand gesture mode is available where the user can open the menu by performing a pinch gesture and navigate using raycasting originating from the hand and pinch gestures to select.
Tapping with a finger on the wrist of the other hand will open the menu in a HoloLens 2. Navigation is then done by tapping directly on the buttons on the interface, similar to a touchscreen (see figure \ref{fig:HL2Start}).

The work by Zielasko et al. \cite{zielasko2019passive} shows that mid-air menus are preferred over menus aligned with the desk. This is also reflected in implementations of current HMDs that present the menu vertically in front of the user.
It is also beneficial to have haptic feedback which is also given in pen and tablet techniques that have shown to be usable for system control.

Many of the works mentioned above focus on special gestures for summoning a system control window when interacting with 3D space around the user. The gestures for summoning are quite large, sometimes using two hands, and they are mostly context free. Information workers, working mostly sitting next to a desk and using small supported hand motions, my not be able to use some of these gestures. Furthermore, it is of interest to use very small gestures that can help the user to summon menus within the context of the data. Gesslein et al. \cite{gesslein2020pen} uses the depth display of HMDs to separate between the planar display of a tablet as the location of the original data and the space above the plane as the place for meta data and menus. The researchers render multiple layers of pie menus in context of 2D spreadsheets (follow the radial menus design of \cite{gebhardt2013extended}) when displayed in VR. The motion of the user's pen above the 2D tablet screen, is used to access stacked radial menus  (see Figure \ref{fig:radialStackedMenu-spreadsheet}).

\subsection{Text Entry}
\label{Section:TextEntry}


\label{TextEntry}
Text entry and editing is a major task of information work, which is a spatial challenge for XR users. The use of near-eye displays can block the view on physical peripherals such as physical keyboards, or touch keyboards, and may interfere with the view of the users own palms and the hand-eye coordination, required for efficient typing.

\subsubsection{Representing Keyboards and Hands}
The best and most popular way to enter text is the full-scale physical keyboard, that has hardly changed in the last century and a half since its introduction. The keyboard supports the user's fingers, and gives haptic feedback, as the user presses each key. 
To leverage on the massive install base of physical keyboards for XR users, there is a need to track it's position and orientation in space and represent it inside the HMD display, as well as the user's hands.

Jiang et al. \cite{jiang2018hikeyb} present a technique called HiKeyb.
The keyboard is being recognized in the depth video and is represented in the virtual space by a corresponding virtual model, while the user's hands are segmented from the depth video and included in the virtual environment as a planar billboard. Users using this technique were measured reaching typing speed of 23.1 words per minute. McGill et al. \cite{mcgill2015dose} used a color video input, and blend the real-time video of the user's hands over the physical keyboard, as a window within the HMD display, showing that users were able to type using this input, although their typing speed was slower compared to their natural typing speed in the real world. Part of this difference might be due to the novelty of the XR environment, and part maybe due to inherent latency or inaccuracies of the system.

Capturing a video of the real-world hands, may not be a solution that fits every application. It's styling may break the immersion of a VR experience and it is limited by the ability to capture the real hands (real world illumination, real world visibility). Researchers have explored other ways of sensing the user's hands actions, from 3D scanning to none at all.

Walker et al. \cite{walker2017efficient} rendered a virtual keyboard in the HMD's virtual environment and did not render the user hands at all. Upon a finger pressing a key on the physical keyboard, the corresponding virtual key lights up. They combined this approach with an auto-correction algorithm, and showed that users could reach a speed of 40 words per minute.  A similar approach by Otte et al. \cite{otte2019evaluating} used touch-sensitive keyboards enabling highlighting touched keys in virtual displays prior to pressing them. They compared fingertip visualization with a touch-sensitive keyboard and found that they are similarly efficient. 

Researchers looked at reconstructing the user's hands geometry and render them in the HMD virtual space as 3D models that follow the user's hand motions and position. Knierim et al. \cite{knierim2018physical} present a system that tracks the physical keyboard and the users hands. They compared different renderings of the hand's models from a realistic rendering, abstract and fingertip rendering both as full opaque objects as well as semi-transparent objects. The performance of experienced typists, which less rely on view of their hands, was not significantly influenced by the visualizations, while inexperienced typists needed some kind of visualization and transparency had no significant influence. For all typists realistic rendering of their hands resulted in higher presence and lower workload.

 \begin{figure}[t]
	\centering 
	\includegraphics[width=1.0\columnwidth]{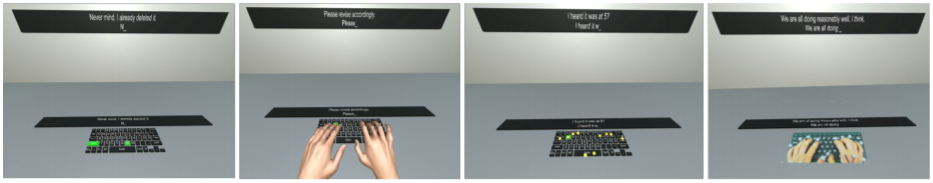}
	\caption{Different representations of the user's hands for typing in a virtual environment \cite{grubert2018effects}.}
	\label{fig:effectsHandRepresentation}
\end{figure}

Grubert et al. \cite{grubert2018effects} compared four different hand representation techniques in the context of text entry: no representation, and video inlay (following the work of McGill et. al), and using tracking of fingertips in 3D space, they rendered two types of 3D models: A minimalist model of the fingertips only, leaving most of the user palm transparent, and a full 3D animated model of the palms. Since the researchers tracked only the fingertips positions, an inverse kinematics technique was used to animate the finger's joints (see Figure \ref{fig:effectsHandRepresentation}). Their study showed no significant differences in typing speed between different renderings. However, using video inlay and fingertip visualization resulted in significantly lower error rates compared to the other two techniques. Surprisingly, using fully animated models increased the error rate of typing, almost as much as using no visualization of the hands at all, probably due to the accumulated effects of small latency and differences between the recovered model and the real hands.
Interesting, the actual speed of typing were not affected by the representation, only the error rate, however the representation can influence subjective measures like presence and workload.

Other works looked at the effect of the keyboard peripheral and the way they are rendered on the text entry quality. The use of XR also opens new possibilities that were not possible in the physical world. For example, since the user do not see their own hands and the keyboard, those may be rendered to the user in new locations, for example closer to the document or the gaze direction of the user.
Dube et al. \cite{dube2020impact} explored the effect of rendering virtual keyboards' keys shapes and choose 3-dimensional square keys. Grubert et al. \cite{grubert2018text} visualized the keys of a physical keyboard in VR and the fingertips of the user.
The researchers also studied the ability to relocate the keyboard and user hands from their physical location to a position in front of the user's gaze. While physical keyboards performance, where the user's fingertips stays lying on the physical keyboard, where not affected by the relocation, soft keyboards, requiring the user's fingertips to raise above the touch surface showed some reduction in speed yet kept a reasonable performance. 

Another way to sense and render the user hands is to use point clouds, generated by depth cameras and LiDARs. While displaying a 3D representation of the hands which might improve hand-eye-coordination, they do not require computational heavy and error-prone processes of recovery like for a full 3D articulated model of the hands. Pham and Stuerzlinger \cite{pham2019hawkey} compared visualizations for typing on a physical keyboard in VR. They compared no VR with the following: no keyboard representation, hand represented as point cloud; keyboard represented as rectangular frame, hand represented as point cloud; virtual model of keyboard, hand represented as point cloud; keyboard and hands shown as video; keyboard and hands represented as point cloud. Authors concluded that the video-see-through is the best option because it is easy to implement and achieves a good entry speed. The point-cloud solution was also found to be competitive, however, it is more complex to implement.

\subsubsection{Mobile Text Entry}
\label{Section:MobileTextEntry}

While most large text entry tasks are still best to be done near a working desk, using a full-size keyboard, the use of wearable HMDs enables the users to enter text also on the go. Researchers looked at using phones as text entry devices for XR users which are common and mobile. One challenge of current phone keyboards is being based on touch, so they require the user's visual sense to guide the fingertips before they touch keys on the phone's screen.  
To overcome this difficulty, Kim et al. \cite{kim2017hovr} use a phone with hovering sensing (sensing finger tips at some distance prior to touching the phone's screen) to visualize both the phone's keyboard and the nearby user's fingertip. Son et al. \cite{son2019improving} used two touch pads with hover function for typing in VR with two thumbs, resulting in a typing speed of 30 WPM. Knierim et al. \cite{knierim2020opportunities} focused on a portable solution and compared the on-screen smartphone keyboard with a desktop-keyboard connected to a smartphone and with a VR-HMD that shows the physical keyboard via video-pass-through. Results indicate a higher input speed in the HMD condition compared to smartphone only, but lower speed compared to a smartphone combined with a physical keyboard. This shows that HMDs with physical keyboards perform better than virtual touchscreen keyboards but worse than physical keyboards without HMDs. 

\subsubsection{Gaze Based Text Entry}
Using XR HMDs opens the possibilities to use new modalities to aid text entry. In recent years, several commercial AR and VR HMDs have introduced integrated eye tracking functionality.
Ahn et al. \cite{ahn2019gaze} and Kumar et al. \cite{kumar2020tagswipe} combined gaze and touch to input text and Rajanna et al. \cite{rajanna2018gaze} combined gaze and a button click. The additional touch modality is used to select a key, and speeding up eye-tracking for text entry, usually using dwell time over a key to confirm inputs. 
Lu et al. \cite{lu2020exploration} explored a hands-free text input technique and compared sensing blinking and neck movements as alternatives to dwell time. Results showed that blinking performed best. Ma et al. \cite{ma2018combining} added a brain-computer interface as a selection mechanism. \\
To date these approaches allow much slower entry speeds compared to physical keyboards and are currently not the first choice for extensive text entry tasks.

\subsubsection{On-surface and Mid-Air Text Entry}
As XR modifies the user senses, it is possible to render virtual keyboards, and turn any physical surface in the environment into a keyboard, enjoying the support and haptic feedback of the surface. 
Richardson et al. \cite{richardson2020decoding} present a technique which combines hand tracking and a language model to decode text from the hand motions using a temporal convolutional network. Participants of a study 
reached a speed of 73 words per minutes (WPM), comparable to using physical keyboards. To achieve such speed the current system was trained for each user for about an hour. 
Fashimpaur et al. \cite{fashimpaur2020pinchtype} also used a language model to disambiguate text entered by pinching with the finger that would normally be used to type a character. This approach, however, achieved a much lower performance (12 WPM).

\begin{figure}[t]
	\centering 
	\includegraphics[width=0.7\columnwidth]{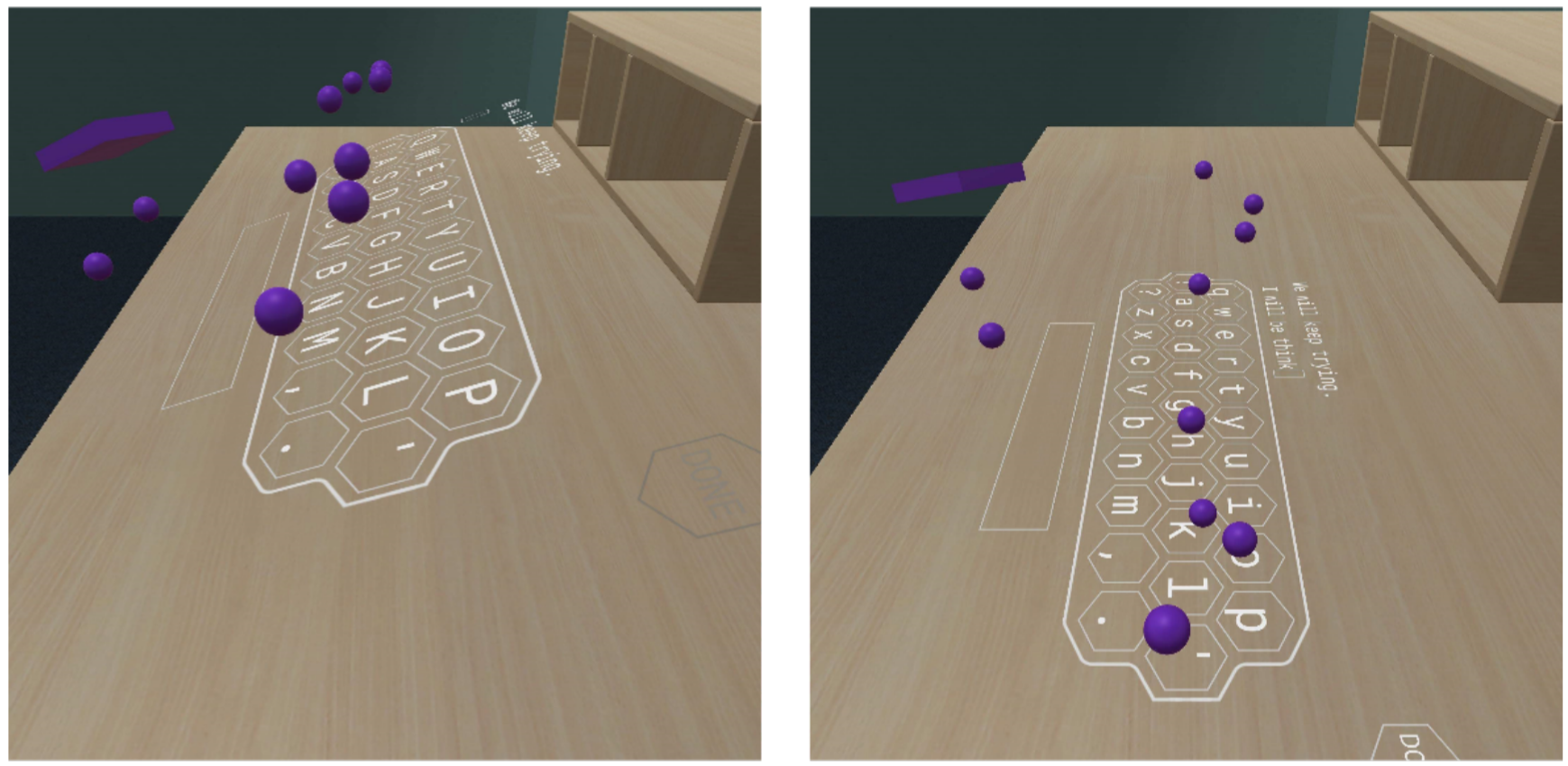}
	\caption{Keyboard and hand visualization as presented by Dudley et al. \cite{dudley2019performance}. On the left, the keyboard is positioned in mid-air. On the right, it is placed on the table. In both cases the fingers are visualized as purple spheres. }
	\label{fig:dudleyPerformance}
\end{figure}

Typing on a virtual keyboard in mid-air lacks haptic feedback. Gupta et al. \cite{gupta2020investigating} explored different tactile feedback techniques for a mid-air keyboard in VR. In their study 
they compared audio-visual feedback to vibrotactile feedback on the fingers, as well as spatialized  and non-spatialized feedback on the wrist. Performance of the four techniques was comparable, but participants preferred tactile feedback. Results also indicated a significantly lower mental demand, frustration and effort for the tactile feedback on fingers. Participants also preferred the spatial feedback on the wrist over the non-spatial. 
Dudley et al. \cite{dudley2019performance} compared typing in mid-air to typing on a physical surface (Figure \ref{fig:dudleyPerformance}), both using only the index finger or all ten fingers and found that users are significantly faster when typing on a surface compared to mid-air. They also reported that participants could not effectively use all ten fingers in the mid-air condition, resulting in a lower speed than the index finger condition.

Other text entry techniques in XR use purely virtual keyboards. This offers higher mobility, because no extra hardware needs to be carried around. Xu et al. \cite{xu2019pointing} and Speicher et al. \cite{speicher2018selection} evaluated different pointing methods for typing on virtual keyboards, including controllers, head and hand pointing. Both concluded that controllers are usually the best choice. Research was also done in evaluating different keyboard layouts for virtual keyboards. For example, using a circular keyboard that can be controlled via head-motion \cite{xu2019ringtext} or a pizza-layout using dual thumbsticks \cite{yu2018pizzatext}. Instead of keyboards, different devices for text entry were explored, like a cube with keys \cite{brun2019keycube}, a ring worn on the index finger \cite{gupta2019rotoswype}, a circular touchpad \cite{jiang2020hipad} or a glove that uses chords to represent letters.
            
However, the performance of such techniques are much lower than for physical keyboards which makes it less suited for longer text entries. Yet they can still be useful in a mobile scenario, when for example writing short messages.

\subsubsection{Using Keyboards beyond Text Entry}
\begin{figure}[t]
	\centering 
	\includegraphics[width=1.0\columnwidth]{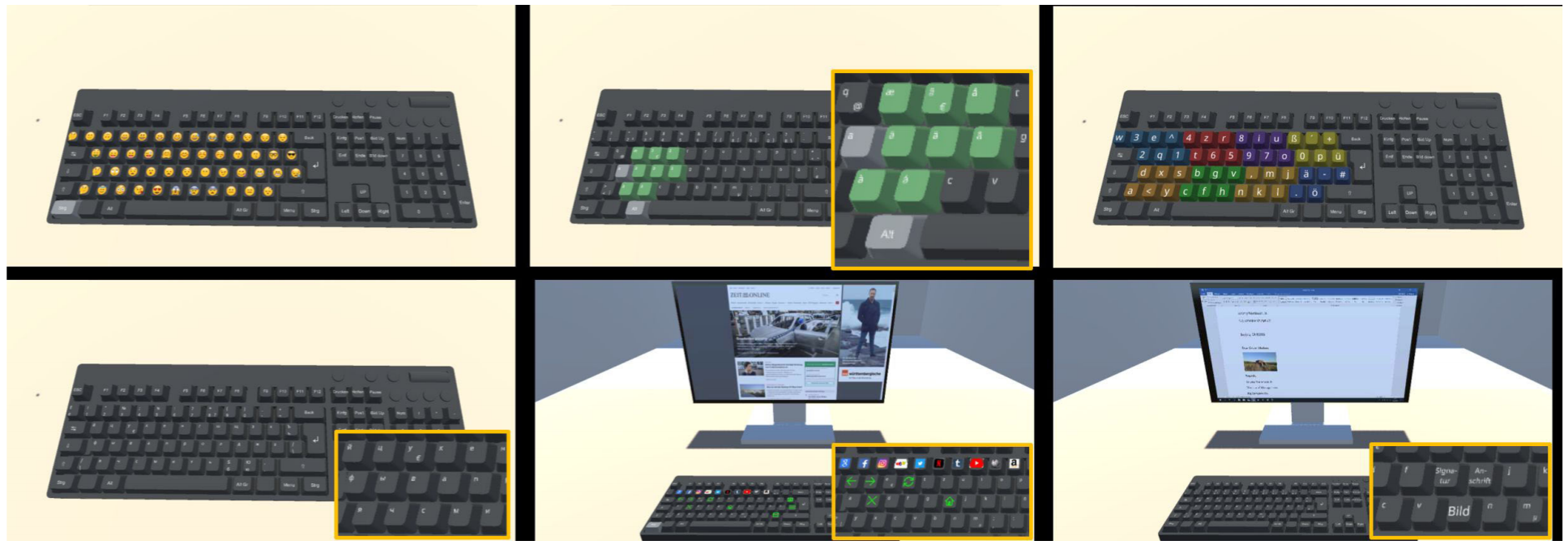}
	\caption{Reconfiguring the layout of a physical keyboard in VR. Top left to bottom right: enabling emoji input; entering special characters; randomized keys for secure password entry; input characters from foreign languages; inlcude browser shortcuts; enable text processing macros.}
	\label{fig:reconviguration}
\end{figure}

Additionally, combining a keyboard with VR can open up new possibilities not available in real world. Schneider et al. \cite{schneider2019reconviguration} explored different input and output mappings enabling different keyboard layouts and functionalities like emoji entry, text processing macros or secure password entry by randomizing the keys, see Figure \ref{fig:reconviguration}.

\section{Collaboration}
\label{Section:collaboration}

XR as immersive technology has a great potential to connect remote coworkers as if they are immersed in the same space. XR's need to track the user's head and hands can be used to animate their avatars, representing their body language. 

On the other hand, XR generates a unique challenge for communication between co-workers by physically occluding the users faces, their gaze and facial expression by the head mounted display. While we are optimistic that in long term, HMDs are going to look like regular prescription glasses and not occlude important parts of the users faces, collaboration software has to bridge this current gap, while using the unique abilities of the immersive displays. Next, we discuss approaches for collaboration in VR and AR.

\subsection{Collaboration in Virtual Reality}
Virtual reality blocks the full view of users and replaces them with a virtual world, generating a special challenge to combine co-workers in the same physical space. Without exact representation of the other users and any geometry in the environment, users may accidentally hit them. 

Programming in a pair, working together at the same time on the same code, has shown several advantages, such as increased knowledge transfer, generation of higher quality code, increased code comprehension and team bonding. Lately, Dominic et al \cite{dominic2020remote}, compared the use of a state-of-the-art remote video sharing to a VR system that represented the avatars of programmers, as well as their keyboard and mice. They found the VR system enables participants to find twice as much bugs and reduce the time needed to solve them. Programmers may have been more focused thanks to reduction of external distractions and they might also have been affected by the immersive feeling of sharing the same space. The authors suggest to further explore the physical and mental demands of collaboration in VR and the effects on productivity and frustration. Sharma et. al \cite{sharma2011workplace}, developed the {\em VirtualOffice} 
system to connect virtually between remote workers along the full work day, including awareness of side-conversations, social happening, and additional opportunities to generate informal communication. The system supported different alert mechanism from text to 3D displays of the virtual office. The authors used Greenhalgh and Benford's concept of an 'aura' attached to each operation to publicized actions and attract coworkers to get involved. The system was designed to use real offices as a base, so it was limited to connected remote workers that work in remote physical offices.  Nguyen et al. \cite{nguyen2017collavr} present CollaVR which enables collaborative reviewing of VR videos using VR HMDs. In addition to watching the video it allows the users to take notes and exchange feedback. In a preliminary study, experts were positive about using VR for such tasks. In the study only up to three people worked together, therefore the authors proposed to do further research on awareness visualizations that are scalable. Additionally, it would be helpful to explore asymmetric hardware setups to include collaborators without access to specific hardware. 
                
While most of the works we discuss in this paper are focused on enabling collaborations in the context of specific tasks, the office environment enables more than a designed environment for performing work tasks. The joining of co-worker together enables generation of informal interactions through chance conversations that are not well supported by current pre-planned teleconferencing. Chow et al. \cite {chow2019challenges} investigated asynchronous interaction in a virtual office environment through virtual recoding and replay of one's action. They also identified and addressed challenges in such a setting such as intrusion of personal space or awareness of actions.

In summary, it has been shown that VR has the ability to increase productivity in collaboration tasks, which is potentially caused by reduced distractions and a feeling of sharing the same space.
A benefit of using VR in knowledge work is to connect workers at remote locations and enable natural, informal interactions that are not possible via teleconferencing systems.

\subsection{Collaboration in Augmented Reality}
Butscher et al. \cite{butscher2018clusters} present ART, a tool for collaboratively analyzing multidimensional data in augmented reality. The visualization is placed on a touch-sensitive tabletop to facilitate existing interaction techniques. They evaluated their design using expert walkthroughs which revealed that AR can support collaboration, a fluid analysis process and immersion in the data. Dong et al. \cite{dong2011collaborative} present another tool for collaboration using augmented reality and tabletops. They developed ARVita where collaborators can observe and interact with simulations in the context of construction planning. 
AR HMDs were also used alongside a tabletop display by Ens et al. \cite{ens2020uplift} to support casual collaborative visual analytics. They evaluated their prototype, Uplift, with expert users and concluded that systems like that have the potential to support collaborative discussions, presenting models or analyses to other high-level stake-holders.

\begin{figure}[t]
	\centering 
	\includegraphics[width=0.8\columnwidth]{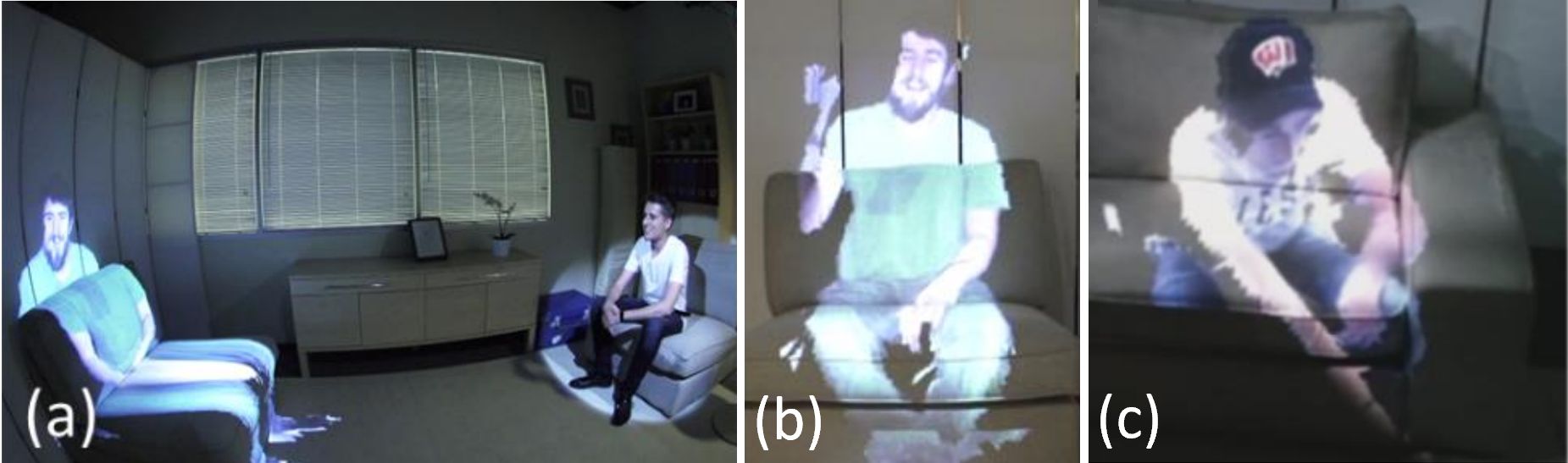}
	\caption{Displaying collaborators in real scale and showing hand gesture helps finish the task faster.}
	\label{fig:Room2Room}
\end{figure}

Pejsa et.al \cite{pejsa2016Room2Room} used projection-based spatial AR to generate a life size view of collaborators, rendered from the point of view of each participant (called Room2Room). Avoiding the use of HMDs, enabled to present the participants in a natural view, life size, using hand gestures that related to the room and using face expressions as part of the communication (see figure \ref{fig:Room2Room}). As the system did not support stereo rendering, the view of the collaborators were set to align with 3D objects in the room, such as sofas or walls. While the presence enabled by Room2Room have not reached that of a physical face to face meeting, the participants reported much more presence than using video conferencing and were able to finish physical arrangement task faster, aided by expressive direction pointing and hand gestures. 
      
Park et al. \cite{park2000lessons} examined a CAVE-based approach for collaboratively visualizing scientific data. They found that users mostly work alone and used localized views to test things without disturbing the overall view and then global views to discuss results with the others. An open question was how the number of participants effects collaborative visualizations. Jing et al. \cite{jing2019snapchart} presented a handheld AR application to support synchronous collaborative immersive analytics tasks such as querying information about content depicted on a physical whiteboard.

Cordeil et al. \cite{cordeil2016immersive} compared HMD and CAVE for collaboratively analyzing network connectivity. They found that HMDs were faster than CAVE in tasks like searching for the shortes path or counting triangles. But no differences in accuracy or communication were found. Additionally they stress the fact that HMDs are less expensive and more readily available.

These works show that AR increases immersion and presence and can support collaborative analyses and presentations. It has also been shown that users value private spaces before sharing their work with others to not clutter the shared space.

\subsection{Hybrid Collaboration}
We have seen examples of collaboration in VR and AR. However, specific devices are not always available to all collaborators. In the following, we present approaches that combine different technologies.

If collaborators do not have access to an XR device or are in situations where using them might be inappropriate, it is useful to enable collaboration between XR and standard applications on desktops or phones. 
Reski et al. \cite{reski2020oh} presented a system for synchronous data exploration using a hybrid interface. It consists of an immersive virtual reality application and a non-immersive web application. They validated the approach in a user study representing a real-world scenario. They concluded that the system fostered experiences of shared discovery and therefore has potential for collaborative data exploration tasks. Future work could have a closer look at the collaborative and communicative behaviour of coworkers inside and outside of VR.

Norman et al. \cite{norman2019impact} presented a study involving collaboration between two local mixed reality users and one remote desktop user. They found that the remote user was more engaged while in the role of a coordinator. Therefore, it is suggested that remote users have specific roles to increase their participation.

Tang et al. \cite{tang2010threescompany} explored the communication channels during a three-way collaborative meeting. The channels include the person space, the reference space using hand shadows, and the shared task-space. Although in this work the setup was not studied in hybrid setting, it can inform how to create hybrid experiences with XR where participants in XR could see virtual hands as reference space and participants without XR could see hand shadows displayed over the task-space.

In some cases, different technologies were also combined to construct new environments.
Cavallo et al. \cite{cavallo2019immersive} presented a co-located collaborative hybrid environment for data exploration. This means they combine high-resolution displays, table projections, augmented and virtual reality HMDs and mobile devices like laptops. Their evaluation results indicate that integrating AR can increase the speed of the analysis process. However, they also state that there are still limiting factors like resolution and field of view of current HMDs.

It can be seen that collaboration is also viable between collaborators using different technologies. In such cases it has found to be useful to assign roles to increase participation. 
Combining different technologies like HMDs, mobile devices and displays provides possibilities to include collocated users with different devices. Yet, the behavior of collaborators in different environments need further research.


\section{Environments}
\label{Section:Environments}
XR has the potential to change the work environment of users, beyond the limitations of their physical environment, and to be designed to reduce stress and increase productivity. We will discuss the importance of including parts of the physical world into the virtual environment ({\em VE}) and how this can be achieved. This also leads to social implications, which can be caused by not being aware of people outside the virtual environment or because bystanders are not aware of what the XR user is doing.

\subsection{Managing Stress and Productivity}
\begin{figure}[t]
	\centering 
	\includegraphics[width=0.8\columnwidth]{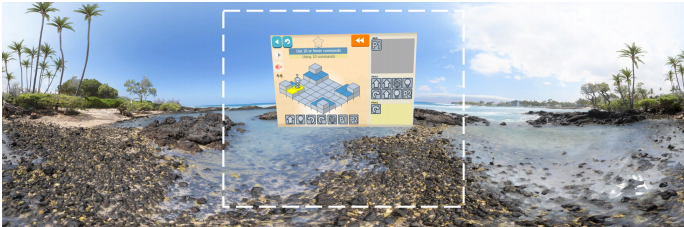}
	\caption{VR replaces the working environment of a worker in an open place to a beach scene, helps the worker stays focused. Image courtesy by Mark Hancock.}
	\label{fig:transportMeAwayRumivova}
\end{figure}

Ruvimova et al. \cite{ruvimova2020transport} evaluated the use of VR to reduce distractions induced by open office environments. They compare four different environments for a task of visual programming: 
a closed office without VR, an opened office without VR, a VR beach environment while the participant was located in a real open office (see figure \ref{fig:transportMeAwayRumivova}), a VR open office environment while the participant was located in a real open office. Results indicated that both the closed physical office and the VR beach virtual environment were equally successful in reducing distraction and inducing flow, and were preferred over the two open office environments. This suggests that VR can be used to stay focused in open offices and the VR environment can be customized to every user's needs. This study focused on single person tasks, the effect of using VR on social interactions between colleagues who share the space is still an open question. 
\cite{anderson2017relaxation, thoondee2017using} showed that their VR application, showing a nature environment, can reduce stress at work and improve mood. 
Pretsch et al. \cite{pretsch2020improving} showed that using VR to experience natural landscapes can significantly reduce stress as perceived by participants and has a significantly higher effect than video streaming similar images.
Valtchanov et al. \cite{valtchanov2010restorative} immersed users after stress-induction task in an explorable VR nature settings and showed that interactive VR nature reduces stress and has positive effect beyond passive VR nature displays.

Li et al. \cite{li2021rear} investigated the influence of physical restraints in a virtual working environment in a study where participants 
did a productivity task in a car while being exposed to different environment. Their findings suggest that users perform better in familiar working environments like an office but prefer secluded unlimited nature environments. They also showed how virtual borders can guide the user to touch the cars interior less.

Lee et al. \cite{lee2019partitioning} used augmented reality as visual separators in open office environments to address visual distractions. The results of their study 
suggests that this technique reduces visual distractions and improves the experience of  shared workspace by enabling users to personalize their environment.

Pavanatto et al. \cite{pavanatto2021we} compared physical monitors with virtual monitors displayed in augmented reality 
and concluded that it is feasible to use virtual monitors for work, yet technically they are still inferior to physical ones. They suggest mixing physical and virtual displays to utilize the enlarged space of the virtual reality and the familiarity of the physical monitors.

All this research shows that XR has the potential to improve the work space of a knowledge worker, by increasing productivity and reduce distractions and stress. 
XR allows the user to create an optimal working environment which can be easily adapted to different situations and requirements. 
VR may currently be ahead of AR in this aspect, as it replaces the entire environment of the user, while AR displays are compared to the real-world quality, yet both can be used to optimize the workplace, by adding additional displays or visual separators.
Multiple studies have shown how stress can be reduced by immersing oneself in a computer-generated nature in VR. It has been shown to outperform 2D displays, and interactions with the display increases immersion and relaxation performance. However, it is not totally clear how it compares to conventional relaxing methods.

\subsection{Including Reality}

Virtual reality replaces the visual and audio sensing of the users with a new virtual environment. As seen above, this new environment may help users to better focus or relax. However, there are objects in the physical environment of the worker that might be of importance for her, like tea cups, desks or co-workers. This sub section looks at research that examines how much of the physical environment users should be aware of while in a virtual environment.

McGill et al. \cite{mcgill2015dose} stated that VR users have problems when interacting with real-world elements and being aware of the real-world environment. They addressed these issues in three studies.
As already mentioned in section \ref{TextEntry}, they showed that typing performance can be significantly improved when enabling a view of reality instead of showing no hands and no keyboard.
In another study they investigated how much reality can be shown and still enable the user to feel present in the virtual environment. They concluded that it was optimal to present the user with reality when the user is currently interacting with real world objects.
Then they studied how this approach could be applied to people instead of objects in social environments to make the user aware of the presence of others.
They concluded that it is important to include some aspects of the physical environment into the virtual. Otherwise, the usability of HMDs in everyday life would be reduced.
Therefore, they propose to blend in relevant parts of reality to preserve immersion while allowing the user to accomplish important actions in the physical environment, like using certain objects, drinking or being aware of other people.

\begin{figure}[t]
	\centering 
	\includegraphics[width=0.6\columnwidth]{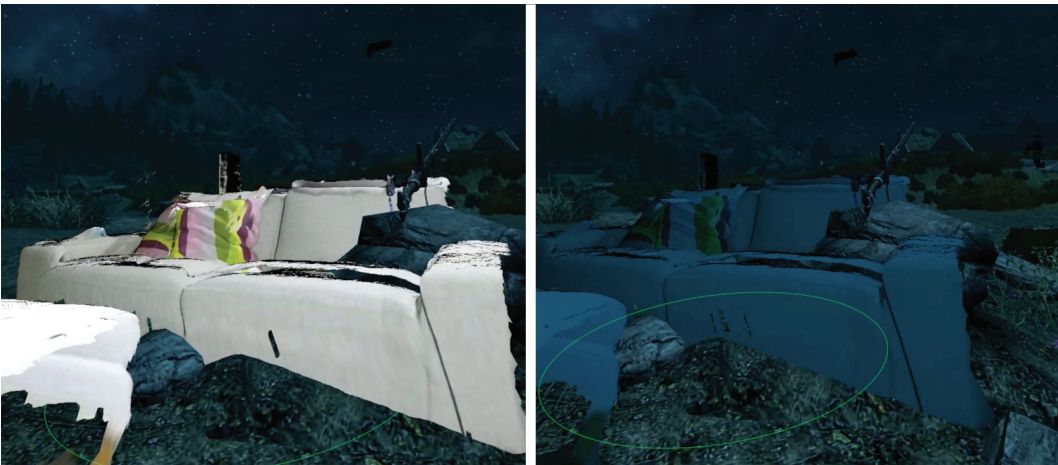}
	\caption{RealityCheck incorporates real-world objects into a VR application (The game Skyrim) (left) and transfer their look to fit the application (right).}
	\label{fig:RealityCheck}
\end{figure}

OneReality \cite{Roo2017One} looks at blending objects over the continuum of virtuality and describe it as a design space. RealityCheck \cite{hartmann2019Reality} takes McGill's approach further by doing real-time 3D reconstruction of the real world that is combined with existing VR applications for the purposes of safety, communication, and interaction. The system allows users to freely move, manipulate, observe, and communicate with people and objects situated in their physical space without losing the sense of immersion or presence inside their VR applications.

O'Hagen et al. \cite{ohagan2020reality} presented "Reality aware VR headsets" that can identify and interpret elements in the physical environment and respond to them. They
evaluated four different notification methods to inform users about the presence of bystanders. Two methods only reported the existence of a bystander and the other two also indicated the position. They detected that some participants were uncomfortable when informed about a bystander with no position information.
In a second study,
they explored dynamic audio adjustment to react to the real world. They either decreased volume to direct attention to a sound in the physical environment or increased volume to block out noise. Results showed that decreasing the volume was effective, but not increasing.

Simeone et al. \cite{simeone2016vr} used a depth camera to detect bystander's positions and shown the information to the user in a "VR motion tracker" which indicated bystanders as a dots in a triangular area that represented the Kinect's field of view. Participants of a preliminary study 
considered it useful and not distracting.

Zielasko et al. \cite{zielasko2019non} compared how substituting a physical desk during a seated task in the virtual environment influences cyber sickness, performance and presence. They did not find a difference between showing and not showing a desk.
However, they argue that showing a desk allows seamless integration of other elements like a keyboard, for which we have seen in section \ref{TextEntry} that it is useful to visualize them.

Knowledge worker use phones regularly. Phones are an important connection to distant coworkers, and may be even more important in a virtual reality, where users are completely isolated from their surroundings. However,  phones, like other physical objects, are usually not included in the virtual environment. Several studies considered enabling using mobile phones in VR.
Desai et al. \cite{desai2017window} presented a system (SDSC) that can detect a smartphone in a video stream, embed it into the virtual environment and fit screenshots sent from the smartphone onto it, allowing VR users to interact with their smartphones while being immersed in the virtual environment.
Alaee et al. \cite{alaee2018user} use depth-based video segmentation to show a smartphone and the user's hands as video pass-through blended in to the virtual environment (NRAV). They compared the technique with SDSC  
and concluded that using NRAV participants could perform several tasks with the same efficiency as when not wearing a HMD and were faster with NRAV than with SDSC. Some users preferred to remove HMDs, but the authors argue that acceptance of the technology will further increase with technological improvements.
Bai et al. \cite{bai2021bringing} presented a technique that brings a virtual representation of the phone and the users hand into VR. Evaluating this technique showed that it successfully connected the real phone with the virtual world, but the experience differed from using a phone in the physical world, with decreased usability caused by hardware and software limitations.

Previous research has shown that it can be very helpful to include parts of the reality in the virtual environment. This is especially true for knowledge workers being immersed for extended periods of time where they need to interact with physical objects like phones or a water bottle or with other people in their surroundings. There are different possibilities to achieve this, similar to including physical keyboards, from virtual replicas to blending in a video stream. For including smartphones, however, the current hardware of HMDs is a limiting factor, as the resolution is not high enough to properly display the content. Regarding collocated people it has been shown to be important to not only indicate their presence but also their position to make the XR user feel comfortable.

\subsection{Social Implications}
When working in everyday environments there are, in many cases, other people around. Therefore, it is important to also have a look at social implications. This includes how the knowledge worker feels when using XR devices, but also how people around the XR user feel and behave.

Bajorunaite et al. \cite{bajorunaite2021virtual} conducted two surveys to explore passenger needs in public transportation that might prevent them from using VR devices. One survey was aimed at an airplane scenario (n=60) and one at ground public transportation, like buses or trains (n=108). For both scenarios, participants expressed concerns about accidental interactions with other passengers and loss of awareness of their surrounding. They suggest to provide cues from the reality and find ways to achieve an engaging experience with less movement. This is in line with the findings presented in the section \ref{Section:Environments}. The results from the surveys also indicate that participants are very conscious of their self-image and how they are perceived by other passengers while using a VR device. They express concerns about being judged because they block out reality. 

George et al. \cite{george2019should} had a closer look at bystanders and explored their ability to identify when HMD users switch tasks by observing their gestures. This could help them find good moments for interruptions. In their study, there was a set of tasks that the HMD user performed (authentication, reading, manipulation, typing, watching video). The bystanders, which were aware of the task set, in the study could identify the task type in 77$\%$ of the time and recognize task switches in 83$\%$ of the time. The authors suggest future work to find out if it has a positive effect on social acceptability when bystanders are able to retrieve meaning from the interactions of the HMD user.

O'Hagan et al. \cite{ohagan2020bystander} studied how comfortable bystanders are in interrupting VR users. Their results indicate that the level of comfort and the acceptability of the interruption strategy is more influenced by the relationship to the VR user than the setting.

Hsieh et al. \cite{hsieh2016designing} presented techniques for socially acceptable text entry, scrolling and  point-and-select realized through hand orientation and finger movement detected by a
sensor-equipped haptic glove which can independently track midair hand gestures. These interaction techniques were considered unobtrusive and socially acceptable.

Research shows that XR users are self concious of the social aspect of their work, and the fear of being judged can limit their use of XR, which may be a subject for future research. It has also been indicated that it could be helpful if bystanders can retrieve some meaning from the behaviour of XR users. This would allow them to better understand them, just like it helps the XR user to be aware of some aspects of the reality to avoid conflicts that disturb bystanders.

We believe that HMDs could be socially acceptable when they become almost not noticeable like regular glasses. Then the issue might be that when the HMDs looks like glasses and become unnoticeable, if the user is sitting at a cafe and doing strange gestures in the air it could look awkward and not acceptable. This is were having subtle indirect gestures from another device such as a tablet could be more socially acceptable (and also require less energy and be more appropriate on crowded space like inside an airplane).
However, some tasks may require large gestures that can not be hidden. Then this is an issue of the population getting used to it, similarly to how we accept people talking to mid air when using an ear piece. If more people use VR HMDs in public it will probably become more accepted, like using a laptop.

In addition, the progress in display resolution could contribute to create HMDs experiences where the main display at the front of the user can contain most of the information in it thus require less head turning on the side or minimize the head turning angle.

Another social implication is the fact that HMDs have cameras which could violate the privacy of others. If we can built HMDs that look like glasses (HoloLens is a step in that direction), we should learn from the privacy lessons from Google Glass.

\section{Extended Exposure}
\label{LOngTerm}

In recent years we have seen a surge in research of knowledge work in XR (AR and VR), showing the possibilities provided by the immersive space.
Different works are aimed at enabling users to work in XR for longer periods of time. For example, one possibility of VR, which hides the user's own body from her sight, includes the ability to use smaller physical motions, while the self-avatar completes a full motion (CoolMoves 2021), reducing fatigue while enabling large interaction spaces. 

However, research on the effect of long-term immersion over full work days, multiple days a week, is very limited. 
Earlier works on eye strain caused by physical displays \cite{stewart1979eyestrain, jaschinski1998preferred} suggest individuals are affected differently. Stewart et al. \cite{stewart1979eyestrain} argued that eye strain results from different factors (i.e. visual, postural, environmental, personal) and that these problems can be solved by considering ergonomics when designing visual displays, suggesting the design of HMDs can be improved to reduce potential problems.

Later research on long periods of using XR has been focusing on the context of manufacturing, therefore we address these works here, even though they do not actually cover knowledge work.
Grubert et al. \cite{grubert2010extended} conducted a study with 19 participants doing an order picking task for 4 hours with and without AR support. Results showed that using AR does not increase the overall objective and subjective strain. However, some participants perceived higher eye discomfort in AR. This was more likely for users with visual deficiencies. They also reported a higher work efficiency in AR compared to non-AR.

Wille et al. \cite{wille2014prolonged} conducted several studies comparing the work with HMDs over several hours with other technologies like tablets or monitors. Objective measures indicated no physiological effects on the visual system and only limited influence of the HMDs weight on the neck muscles which contrasts with the subjective ratings. The authors speculate that the unfamiliar technology influences the subjective ratings. 
Funk et al. \cite{funk2017working} studied an industrial assembly task with instructions projected directly on work piece, and using a depth camera (Kinect v2) to verify assembly correctness. All participants used the system for at least 3 full working days. 
The results showed that the instructions were helpful for untrained workers, however it also slowed down the performance and increased cognitive load, especially for expert workers. 

Steinicke and Bruder \cite{steinicke2014self} conducted a 24-hour self-experiment in virtual reality (using Oculus Rift DK1 HMD) with one participant, who worked, ate, slept and entertained himself with music or movies. They reported higher simulator sickness after periods involving many movements, and lower values when resting. The participant reported limitations of the HMD due to latency when moving and a limited resolution when working. It was also reported that the participant was sometimes confused about whether he was in VR or a real environment and that the perceived accommodation seemed to vary after a few hours. 

\cite{nordahl201912} had two participants using VR HMDs for 12 hours. They used Oculus Rift CV1 HMD for 6 hours and the HTC Vive HMD for the rest. They only took the HMDs off for switching them after 6 hours. While in VR the participants used different applications. The reported results indicate that simulator sickness symptoms were mild with a peak after 7 hours that is difficult to explain because the experiment was not fully controlled.

The most extensive research on long-term immersion has been conducted by Guo et al. \cite{guo2019mixed, guo2019evaluation, guo2020exploring} and Shen et al. \cite{shen2019mental}. They applied Maslows Hierarchy of Needs to guide the design of a VR office. First, they conducted a short-term study (n=16) \cite{guo2019evaluation} using a text input and image processing task which was done in the virtual and a physical environment. They concluded that the designed VR office was comfortable and efficient and therefore, in a next step, used it for a long-term study.

In the long-term experiment \cite{guo2019mixed} 27 participants were in the virtual and physical office for 8 hours each, doing knowledge worker tasks like document correction, keyword searching, text input and image classification. 
They compared the results with the short-term study and concluded that physiological needs like drinking, belongingness needs like communication, temporal presence and self-presence are important for long term immersion but can be ignored for short-term. Safety needs and emotional needs on the other hand must be met in both conditions. 

This 8-hour experiment was also used to investigate mental fatigue differences between the virtual and physical work spaces \cite{shen2019mental}.
Participants performed a psycho-motor vigilance task (PVT) 6 times during the experiment, and results showed that there were significantly less PVT lapses in the physical environment and the reaction times were slower in VR, indicating a higher mental fatigue in VR.
The authors propose two explanations. Either the additional visual information processing in VR occupies more attention resources, or that VR can increase attention of participants more effectively which lets them allocate higher attention resources.

The same experimental setup was also used to explore the difference in visual discomfort between working in VR and physical environment \cite{guo2020exploring}.
The results showed that subjective visual fatigue, pupil size and accommodation response changes with time in both conditions. They also detected a gender difference suggesting that female participants suffer more from visual fatigue in VR than male participants. However, this could be caused by male participants in the sample having more experience with VR than the female participants.
There was also no significant difference in nausea which can be explained by the static content. The authors recorded no significant eye strain difference between VR and the physical environment, yet focus difficulty was significantly higher in VR which might be caused by the accommodation vergence conflict.
These studies showed only mild simulator sickness symptoms  
which can be explained by displaying relatively static content in knowledge worker tasks.
A large part of physical discomfort is due to the weight and form factor of HMDs, which will hopefully become much more comfortable in the near future.

Shen et al. \cite{shen2019mental} reported higher mental fatigue in VR compared to a physical environment. Additionally, in VR the reaction time significantly increased over time but not in the physical environment.
The authors propose two explanations. Either the additional visual information processing in VR occupies more attention resources, or VR can increase attention of participants more effectively which lets them allocate higher attention resources.

From those studies it can be concluded that simulator sickness symptoms are rather mild \cite{nordahl201912, guo2020exploring} which can be explained by the relatively static content in knowledge worker tasks.
A large part of physical discomfort is clearly due to the weight and form factor of HMDs. As HMD hardware is being improved, mainly on resolution, frame rate, weight and even dynamic focal distance we expect the VR experience become more comparable to physical environment one, and more such long-term studies 
will be required.
Existing research points to important directions when designing applications for long-term use like latency \cite{steinicke2014self}, physiological needs, emotional needs, safety, presence and belongingness \cite{guo2019mixed}. 
Experience can be heavily influenced by personal traits like impaired vision 
, or XR past experience.
Therefore, longer studies are needed, so the participants get accustomed to the new technology which will make it much more comparable to a physical environment.

    


\section{Applications}
\label{Section:Applications}

This section will have a closer look at examples of applications that support knowledge work in XR.

There are several application examples involving data analysis.
Zielasko et al. \cite{zielasko2017remain} discussed the potentials and benefits of using VR HMDs in a desktop configuration, i.e. when being seated in front of a table. They described scenarios and use cases in the domain of data analysis. 
A similar proposal was made by Wagner et al. \cite{wagner2018virtualdesk, wagner2019comfortable}. On both instances, users interacted with data using mid-air interaction using their hands.

Another application scenario that benefits from XR is programming.
Elliot et al. \cite{elliott2015virtual} presented concepts for supporting software engineering tasks such as live coding and code reviews using VR HMDs in conjunction with mouse and keyboard interaction. 
Dominic et al. \cite{dominic2020remote} compared remote pair programming and code comprehension in VR against a standard shared display. In their study (n=40) the average time to solve a bug was lower in VR and they solved nearly twice as many. Authors explain this difference through reduced distractions in VR and a sense of being collocated with the collaborator. 

XR has also been explored for creative tasks, for example Nguyen et al. \cite{nguyen2017vremiere} proposed a video editing system for VR.

As the resolution, size and weight of HMDs evolves to more consumer oriented glasses-like form factor, HMDs becomes an attractive alternative to physical monitors \cite{pavanatto2021we}, being more mobile, large and private. 
In recent years we see research that looks at ways that common 2D office taks may be enhanced when done via near-eye displays.

\begin{figure}[t]
	\centering 
	\includegraphics[width=1.0\columnwidth]{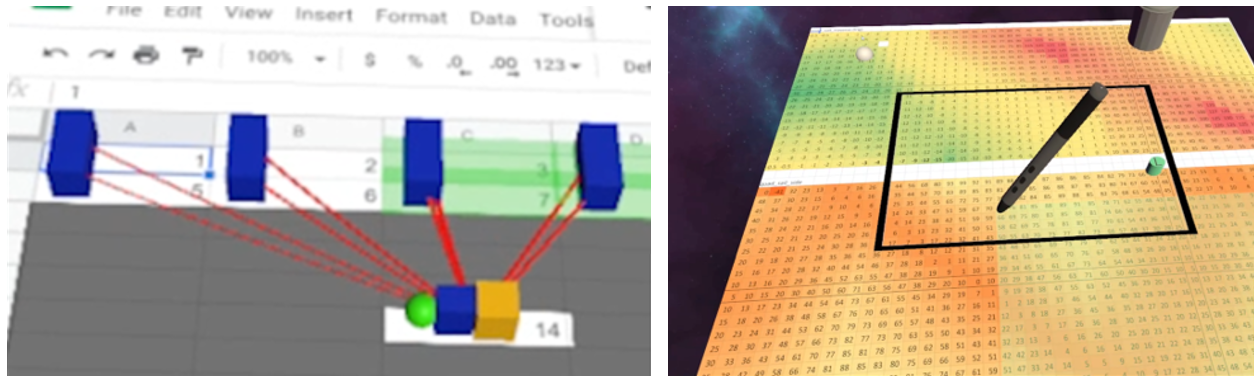}
	\caption{The left image shows relations between cells. A function is applied to the cells at the top which results in the value shown in the bottom cell. The right image shows how the spreadsheet is visualized beyond the boundaries of the tablet used for interaction which is outlined in black. }
	\label{fig:spreadsheetApplication}
\end{figure}

Gesslein et al. \cite{gesslein2020pen} enhanced a spreadsheet application in VR which enabled new functionalities not possible in 2D, like visualizing relations between cells above the screen or extending large sheets beyond what is possible on physical screens (see Figure \ref{fig:spreadsheetApplication}).
O'Brien et al. \cite{o2019wikipedia} presented a virtual environment for browsing wikipedia content.
Biener et al. \cite{biener2020breaking} propose many different applications that leverage the possibilities of VR in the context of mobile knowledge workers. For example, they present an array of layered screen arranged around the user that expands the users small portable display and an application that visualizes 3D data on top of a touchscreen device (Figure \ref{fig:breakingTheScreen}).
Dengel et al. \cite{dengel2006human} and Deller \cite{deller2008managing} presented applications for working with documents on a stereoscopic display

There are also commercially available solutions that support knowledge work in virtual reality, such as Oculus Infinite Office \cite{oculus2021} where the user can open virtual browser windows and interact with them using a controller, gestures or a physical keyboard which is included in the virtual environment as a 3D model and a video pass-through visualization of the hands while typing.
An example for AR would be spatial.io \cite{SpatialIo2021} which allows collaborative meetings by representing remote coworkers as avatars


\section{Challenges}

Informed by the presented review of literature, we see several challenges that need to be addressed in the future to make every day virtual and augmented reality a compelling experience that every knowledge worker wants to use and possibly prefer over a real office. This includes evolving HMDs, improving the overall user experience in XR, and embracing other devices and peripherals around the user.

\subsection{Challenges in the Next Generation of HMDs}
The technology of consumer available XR HMDS has progressed by leap and bounds in recent years. It is possible nowadays to see HMDs with a resolution of 5K, and refresh rates of up to 144Hz.
However, for being good monitor replacements for information worker there is a need to display text at quality that is comparable to 2D monitors. First, such HMDs needs to be small and light as glasses to be able to be wore for a whole day. Second the current lenses that are used in commercial HMDs are simple Fresnel lenses that reduce the display quality further away from the center of the display, many times to a level where text is unreadable. Multi layered lenses and other optic solutions that are able to generate sharper displays, may be too expensive for current HMDs that are used mostly for entertainment.
The combination of eye tracking and changeable focal distance of the display or light field display, can reduce fatigue originated from inaccurate convergence/focus rendering.


When using XR for longer periods, there is more importance for sensing around the headset, from the structure of the environment around the user, enabling the user to move around and use different physical resources, to interacting with co-workers, and update of the XR space and applications according to the changing conditions around the worker.
For example, precise and responsive hand tracking is important for pointing or text entry. Detection of peripherals such as physical keyboard (e.g.  supported by Oculus Quest 2) enable representation of them in the HMD display, where XR applications can use them and even modify them for their needs \cite{schneider2019reconviguration}. Better environment sensing via cameras and depth sensors enables users to move naturally in virtual offices while being in a uncontrolled dynamic environment and avoiding obstacles \cite{cheng2019vroamer,yang2019dream}.

The form factor and design of the HMDs also impacts its comfort and social acceptability. It has been seen in section \ref{LOngTerm} that a great part of users discomfort during the use of HMDs is the weight. As manufacturers are working on lighter and thinner HMDs, we hope they will be comfortable for extended use. Glasses-like form factors may also help with with social acceptability as it is less disconnecting workers from their environment, and make them easier to communicate and collaborate.

\subsection{Challenges in Future XR Experience for Knowledge Worker}
There are also several design challenges that need to be addressed to make XR experience appealing to knowledge workers.

One of the advantages of using XR for work, is the independence of the physical conditions in the user's environment. When workers travel, or as they work from home, they might find themselves in physical environments that are less than optimal. In many cases, users need to limit themselves to small spaces, such as an airplane seat, while they are expected to work at their best capability. Working applications should be aware of the user's limitations and display a virtual working environment that is fully controlled by the users given their limited input space. While current XR applications require users to prepare an empty physical environments to enable uninterrupted and safe working spaces, future applications will have to be flexible and adaptable to a variety and dynamically changing environments \cite{gal2014flare, yang2019dream}.

Given the possibly limited input space, and the need to reduce fatigue of the worker, new mappings will enable users to execute small motions, mostly supported, according to their or the environments limitations, and have them mapped to large effects in the virtual space \cite{ahuja2021cool,wentzel2020improving}. 
The environment around the user may exhibit the change of tasks, and enable fast switching of context. It may help relax the user when it is needed, and keep her aware of the presence of other people in her work vicinity both physical and remote. 
The representation of the user should bridge the difference in presence between local and remote users, for example, by representing physical affordances such as white boards, meeting rooms, etc. in the virtual domain too. workers without XR displays need also to be aware of the presence of remote users, with range of methods from large displays and representation robots to projections \cite{pejsa2016Room2Room}.

\subsection{Embracing other Devices and Peripherals}
Previous studies indicated that it is beneficial if the system also embraces other devices and peripherals close by.


For example, physical keyboards should be seamlessly integrated in the experience and possibly augmented. Because text input is a crucial part of knowledge work, the typing speed should be as good, or maybe even better, if effective ways to augment the keyboard are found, then in the real world.

Another useful device would be a stylus that is tracked with no delay and the system knows exactly when it is touching or raised from any surface which could be a digitizer or just an table or wall. Tracking touch down and touch up events of styli or fingers on a passive surface is very challenging because the system needs to see exactly when the touch occurs. A stylus could send messages when touching a surface, however, this would not work for fingers. Yet vision or audio-based solutions could also be very compelling.

Finally, all devices such as tablets or smartphones that are close by should get seamlessly integrated into the experience and be used according to their properties to enrich the XR experience.

\section{Summary and Conclusion}
In this chapter we have presented an overview of interaction techniques that can support knowledge workers in XR.

Techniques for interacting with 2D content in 3D included controllers, gloves, hand tracking, tablets, eye gaze, pens and more. 
Studies have shown that it can be beneficial to use indirect input which reduces fatigue while preserving performance.
Researchers are already working on socially acceptable techniques and future research should also keep that in mind, because this is an important aspect in everyday environments.
It has also been found that touch surfaces perform better than mid-air which is in line with the findings from exploring text entry techniques.
In that context it has also been shown that users prefer tactile feedback, for example provided by a physical surface, over mid-air typing.
For text entry it is very popular to include physical keyboards in the virtual environment and different visualizations for keyboards and hands have been explored like video-pass-through, point clouds, virtual models and abstract representations.
It has been shown that the type of visualization influenced workload and presence, but not performance, as long as there is some kind of visualization.
Typing on any physical surface has also been shown to be effective when combined with language models and convolutional networks.
However, currently such techniques need to be calibrated for each user which decreases usability.
Additionally, many other techniques and devices have been explored including eye-gaze-typing and phones, usually resulting in much lower entry speeds.
While this could be negligible for some use cases it is an important factor for knowledge workers.

We also presented approaches on how collaboration can be supported in XR for tasks like pair-programming, video reviewing or exploring information.
It has been shown that such systems can perform better than standard video tools and increase presence.
A problem that arises is how the presented systems can be scaled to allow collaboration of more people, as usually the studies only involved two or three users.
We also presented solutions for including users without XR devices, for example via a web application. However, it is important to design such applications in a way that engage participation from the non-VR user.

Research on virtual environments have shown that VR can be used to reduce distractions in office environments, because the virtual environment can be customized to the user's needs.
VR has also been found to be able to reduce stress, for example by showing natural scenes and that users prefer such scenes even though they perform better in familiar environments like offices.
Future work could investigate which environments are suited best for which situations.
Research also suggests to include some parts of the physical environment in the virtual environment.
This can be done by blending in relevant parts, like tables or objects on the table. They can be included as video stream or as virtual replicas while preseving immersion.
Similar to keyboards this can also be done for phones. However, the limited resolution of today's HMDs makes it hard to use such small devices.
Besides objects, it is also helpful to include information about bystanders, like presence or position.
This can help the VR-user to avoid undesired bahavior like accidental interactions.
However, further research is needed on how VR-user and bystanders feel and how they are affected.

We have shown that some research has been conducted on the effects of extended usage of XR devices such as VR and AR HMDs.
Unsurprisingly, a main issue is physical discomfort caused by weight and form factor of HMDs which will hopefully improve as technology advances.
Studies also suggest that the experience can be influenced by factors like latency, safety, presence, impaired vision or experience.
To date the longest study was 24 hours, therefore more research is needed on longer even periods of time.

Finally, we presented some applications that show the value of using XR for knowledge work.
Several works show that XR can be beneficial in the area of data analysis, including spreadsheet applications.
XR has also been shown to support collaborative programming and it can be utilized to increase limited screen space in mobile scenarios.

%
%

%
%
\bibliographystyle{abbrv-doi}
\bibliography{template}
%

%
%
%
%
%
%
%


\end{document}